# Quality enhancement with simultaneous suppression in excess noise of longer-lived states in non-uniformly pumped optical microcavities

Arnab Laha, and Somnath Ghosh*
*Institute of Radio Physics and Electronics, University of Calcutta, Kolkata-700009, India*

We study the formation of longer lived states via internal coupling near a special avoided resonance crossing between the interacting states in an open optical system. Away from $\mathcal{PT}$ symmetry limit, we discuss unconventional cavity resonance states created by a spatially varying gain/loss profile. Near certain avoided-resonant crossings ($ARC$), the parity symmetries of the chosen coupled states dictate the destructive interference resulting high $Q$, while avoided crossing in the imaginary part of the poles embraces the longer lived state. Now, via $S$-matrix formalism, we have numerically demonstrated that with suitably chosen system openness and coupling strengths, the excess noise generation among the interacting non-orthogonal states (calculated as *Petermann* factor $K$) can be suppressed close to the ideal value 1 with simultaneous order of magnitude enhancement in $Q$-factor of the longer lived state.

## I. INTRODUCTION

Intrinsically it is extremely challenging to confine light in a small space over a length scale comparable to the wavelength of light. Enabling strong light/photon confinement for a long time, measured as photon lifetime in an optical microcavity enhances light-matter interaction significantly. This phenomenon plays an important role in nonlinear optical processing including isolation [1–3], quantum information [4–6] etc; and opens up possibilities for development of indispensable optoelectronic devices for wide range of on-chip/integrated photonic applications. To serve this purpose to confine light indefinitely in a cavity, it should be ideally lossless. Any deviation from this ideality of a fabricated optical microcavity is characterized by a parameter that quantifies the degree of confinement, known as quality factor ($Q$-factor) and it is proportional to the confinement time in units of the optical period [7]. Light matter interaction in such leaky/open cavity can be well explained by using non-hermitian formulation of quantum mechanics. According to which, due to interaction between light wave and the surrounding environment via continuous exchange of energies, the amount of energy being carried by the individual states is not conserved and vary in time [8]. Hence, the energy eigen-values of quantum states in such systems can be written as complex-valued numbers i.e. $E_j = \Re(E_j) \pm \Im(E_j)$. Here $\Im(E_j)$ determines the ability of confinement in terms of lifetime i.e. $(\tau_L)_j \propto 1/|\Im(E_j)|$ and $Q$ factors are given by $\Re(E_j)/2 |\Im(E_j)|$. Recent developments in the fabrication technology for growth of optical micro-resonators with high precision and control have resulted in, extensive contemporary research interests in the enhancement of $Q$-value by forming a longer lived state (i.e. smaller $\Im(E_j)$) which has already been explored via the existing routes of photonic crystal defect modes [9], metallic nono-cavities [10–12], and semiconductor nanowires [13] in the context of resonance interactions. Tremendous efforts have been put forward to develop an efficient optical cavity to enhance the critical parameter $Q/V$ ratio to increase the mode density per unit volume and minimization of associated loss issues while keeping the require ultra-high $Q$-values.

In this direction, lately *proximity resonances* have attracted lot of attention along with new physical insights for enhancement of $Q$-factor. This phenomenon was first reported by Tolstoy *et al.* where it was coined as *super-resonance* and identified as a purely wave based phenomena [14]. The electromagnetic resonances of two individual scatterers placed within single wavelength can be coupled via evanescent field and form proximity resonances. This can be easily understood by the alternate route of quantum scattering theory. Accordingly, during scattering phenomenon between two almost identical scatterers, near resonance the individual scattering cross-sections become extremely large compared to the physical size of the scatterers which can be placed sufficiently close together. Subsequently, due to presence of an in-phase symmetric mode, the field amplitude become twice than that of a single one; and the radiated power become four times which is twice the power for incoherent consideration of two scatterers [15]. The proximity resonance may introduce significantly higher $Q$-value comparably that of the resonance of a single isolated scatterer *i.e.* when two scatterers are far apart, nevertheless lower than that of the scatterer having same length and a cross section equal to the total area of two previously considered isolated scatterers [16]. In an open system each resonance is coupled individually to the continuum. Now in such system two resonances may also be coupled internally (almost like a hermitian coupling of two states in a closed system, but the only difference is that each state is individually coupled to the continuum) and/or externally (coupling through the continuum which is fully non-Hermitian). For strong coupling, the resonant frequencies anti-cross and the wave functions exchange; whereas for the case of weak coupling, the resonant frequencies cross and there is a slight mixing of wave functions but no exchange of identity. Its reported previously that the strong external coupling as well as weak external coupling lead to a great enhancement of lifetime and can form long-lived states [16, 17]. *Avoided resonance crossing* (*ARC*) plays a key role in this context.

*Electronic address: somiit@rediffmail.com

*ARC* is a generic phenomenon which was originally studied in chemistry by Neumann and Wigner in the context of transition of electronic states and later it is widely studied for diatomic and polyatomic molecules [18–20]. Electronic states of a diatomic molecule do not cross, unless permitted by symmetry criteria resulted in an *ARC*; whereas the states having different symmetry are allowed to cross and the states must be degenerate at the point of crossing [21]. In the context of standard quantum mechanics, a hermitian matrix has been considered to represent a system that depends on some real continuous parameters. Eigen-values of this matrix, corresponding to the symmetric eigen-states, cant cross *i.e.* there should be no two or more equal eigen-values, resulting in a phenomenon known as avoided crossing. In other words, the curves representing the two energy eigen-values, as function of a real parameter say $\Delta$, when approach a crossing, owing to stronger coupling and symmetry restriction, instead repel each other [17]. Recently, the avoided resonance crossings (*ARCs*) in open /non-ideal systems have attracted enormous research interest along with potential technological applications. Such phenomenon when associated with a transition/branch point [22], is of particular interest and has received considerable attention in atoms or molecules in magnetic field [23], microwave cavities [24], and optical micro-cavities [15, 17].

In this paper we choose internal coupling of resonances in a $1D$ two port open optical micro-cavity to form long-lived state. Under certain condition in a strong coupling regime in the gain/loss cavity, a frequency dependent avoided crossing between two internally coupled resonances results in extremely high $Q$-factor of the desired resonance state. Simultaneous suppression in excess noise factor as much as close to 1.86 is achieved for potential stable performance of the high-$Q$ cavity.

## II. MATHEMATICAL PICTURE: AVOIDED RESONANCE CROSSING AND EXCESS NOISE

Coupling of two resonances can be fully explained in terms of a $2 \times 2$ Hamiltonian matrix [17] which has a fundamental importance in quantum mechanics, because it embodies simplification of many of physically realizable systems. Quantum mechanically this matrix indicates a two level quantum system with energy eigen-values $E_1$ and $E_2$ and the system is subjected to an external perturbation. For the sake of simplicity, the perturbation is considered as only off-diagonal elements i.e.

$$H = \begin{pmatrix} E_1 & 0 \\ 0 & E_2 \end{pmatrix} + \begin{pmatrix} 0 & V \\ W & 0 \end{pmatrix} = \begin{pmatrix} E_1 & V \\ W & E_2 \end{pmatrix} \quad (1)$$

Here $E_1$ and $E_2$ are complex energies of the uncoupled system; $V$ and $W$ are coupling terms.

The eigen-states of the coupled system can be obtained by diagonalization of the matrix, and eigen values are given by

$$E_{\pm}(\Delta) = \frac{E_1 + E_2}{2} \pm \sqrt{\frac{(E_1 - E_2)^2}{4} + VW} \quad (2)$$

The case $V = W^\star$ ($VW$ is real and positive) represents internal coupling whereas $V \neq W^\star$ ($VW$ is complex quantity) permits external coupling. In this paper we deliberately consider internal coupling only. However, our design cavity would preferably operate near a transition regime from weak to strong coupling strengths. If we plot $E_+$ or $E_-$ with $\Delta E$ i.e.($E_1 - E_2$), two branches of hyperbola should be obtained. Analyzing these curves it can be concluded that the repulsion between the energy levels entirely depends on the coupling parameters $V$ and $W$. Mathematically, if $W$ is set to zero then $E_+$ should be equal to $E_-$ at $\Delta E = 0$ *i.e.* the energy levels cross. Thus the level crossing should be avoided only by introducing a perturbation. Physically the full features of typical *ARCs* in open or dissipative system are captured in matrix (1) in such a way that far away from the *ARC*s we can still ignore the off-diagonal terms and safely follow the *ansatz* $| E_1 - E_2 |^2 \gg VW$. Now keeping the restriction as $V = W^\star$ permits internal coupling which allows two different kinds of *ARCs* [18]. When $VW > | \Im(E_2) - \Im(E_1) | /2$ then the coupling is strong enough to cause an avoided crossing in the real parts and crossing in imaginary parts of the energies. When $VW < | \Im(E_2) - \Im(E_1) | /2$ then the coupling is weak to cause an avoided crossing in the imaginary parts and crossing in real parts of the energies. The later one has been exploited to design optical microcavities with unidirectional light emission from long-lived states [25].

The *Petermann* $K$-factor, introduced by K. Petermann [26] in 1979 for gain-guided lasers, makes a contradiction with classical formula of $K$-factor (or excess noise factor), proposed by Schawlow and Townes [27] in 1958 in the context of quantum limited line-width of a laser cavity. Even the Schawlow-Townes formula was set in favour of quantum mechanics; Petermanns result was apparently a violation of fundamental principles. Later, Haus and Kawakami [28] had tried to resolve above contradiction by pointing out the correlation between the noises terms associated with different propagating modes in a loss- or gain-guided cavity, which was eventually in a good agreement with the fundamental principles of quantum mechanics. The excess noise factor is not only just a property of gain- or loss-guided lasers but also of any resonance cavity with non-orthogonal transverse eigen-modes [29]. Actually, it is the fact that the transverse modes which are far from being orthogonal, lead to the appearance of a large excess noise factor. Hence, excess noise factor can be studied in unstable resonators where many non-orthogonal modes are supported naturally or in non-hermitian stable resonators where the supported modes are inherently non-orthogonal along with other geometrically deformed/ open interacting cavities [30, 31]. For example, most of the geometrically stable cavity resonators, where the transverse eigen-modes are orthogonal, dont show a noise factor unless the supported polarization modes can be made non-orthogonal [31]. Such types of resonators with supported non-orthogonal modes are simpler than an unstable resonator with many transverse modes for studying excess quantum noise. The noise factor in such non-hermitian stable resonator/ cavity has been discussed and measured previously [32, 33]. However, our interest is to study excess noise factor in a two-port open cavity where mode-mode interaction is being explored to achieve ultra-high Q-factor from the view point of *proximity resonances* unlike the case of a stable cavity [34]. It is now established that for either cases of polarization-mode and transverse-mode, the $K$-factor can be described convincingly by the mode non-orthogonality theory [29]. However this formal connection between non-orthogonal modes to excess noise, does not conclusively state the physical origin of the excess noise factor.

This noise factor has been extensively studied theoretically as well as experimentally in laser cavities [26, 29, 35–37]. If $\Psi_n$ represents the eigen function corresponding to the $n^{th}$ supported mode then the *Petermann* factor can be expressed as $K_n = 1/|<\Psi_n^\star | \Psi_n>|^2$ [30], which gives the unit value of noise factor for ideal orthogonal modes. In this picture, the quantum limited line width ($\delta\omega$) of radiation from the lasing mode can be expressed in terms of resonance width ($\Gamma$), noise factor ($K$) and the output power ($I$) [26]. Its given by- $\delta\omega = K(\Gamma^2/2I)$ which is larger than the Schawlow-Townes classical value ($\Gamma^2/2I$) [27] , i.e. the quantum-limited line width of a laser cavity is enhanced above the classical value by the *Petermann* factor-$K$ and which is exactly due to the non-orthogonality of the cavity modes.

### III. TWO PORT OPTICAL CAVITY AND S-MATRIX FORMULATION

Recently, unconventional laser-absorber modes, which are created by spatially varying gain/loss profile have been widely studied in the context of $\mathcal{PT}$ symmetry cavity [38–40]. Nonetheless, in this section away from the $\mathcal{PT}$ symmetry limit, we study unconventional cavity resonances. Cavity is accompanied by an avoided crossing of the poles of the corresponding $S$ matrix, which are analogous to the complex eigen-values of the associated non-hermitian Hamiltonian of the cavity. For brevity, we consider a partially pumped $1D$ two port open coupled optical cavity of Fabry-Perot-type, as shown in FIG. 1 ,following the coupling restriction in the non-hermitian regime. Choice of such Fabry-Perot geometry was driven by the fact that we can implement the output coupling along the axis, which is otherwise not possible in toroidal resonators without proper phase matching. However the results reported here are generic and is extendable to any other resonator geometries. The cavity occupies the region $0 \leq x \leq L$, where the gain (denoted as co-efficient $\gamma$) is introduced in the region $0 \leq x \leq L_G$ with refractive index $n_G$ and a loss $\tau\gamma$ ($\tau$ is the fractional loss co-efficient) is introduced in the region $L_G \leq x \leq L$ with refractive index $n_L$. Thus the system is designed in such a way that if certain amount of gain is added then proportionate loss is introduced simultaneously and dictated by $\tau$. While designing our system we choose $L$ and $L_G$ to be $10\mu m$ and $5\mu m$ respectively; the value of $\tau$ is taken as $0.274$. The particular loss to gain ratio is chosen to be near a special transition point of two different categories of $ARC$s supported by the cavity while gradually increasing the $\tau$ from 0 to 1. This variation of $\tau$ necessarily governs system openness and coupling strengths between the states. Motivation behind this specific choice of $\tau$ value is comprehensively explained in the next section. Introducing $n_R$ as real part of the background refractive index, $n_L$ and $n_G$ can be written in terms of gain and loss as follows:

$$n_G = n_R - i\gamma$$
$$n_L = n_R + i\tau\gamma$$
(3)

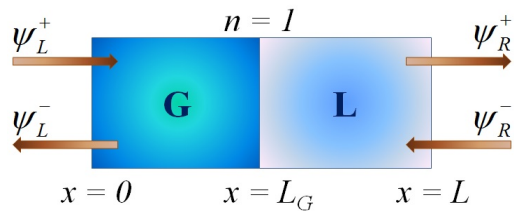

FIG. 1: (Color online) Schematic diagram of a two port open interacting optical cavity that occupies the region $0 \leq x \leq L$. Gain and loss is introduced in the left and right sides of $x = L_G$. Inside the cavity, uniform background refractive index is chosen to be 1.5. $A$ and $D$ are the complex amplitude of incident wave corresponds to the wave function $\Psi_L^+$ and $\Psi_R^-$ whereas $B$ and $C$ are the complex amplitude of scattered wave corresponds to the wave function $\Psi_L^-$ and $\Psi_R^+$ respectively.

Here $n_R$ is chosen as $1.5$ for fabrication feasibility of practical devices. The $S$-matrix of such specially designed cavity can be defined as

$$\begin{bmatrix} B \\ C \end{bmatrix} = S(n(x), \omega) \begin{bmatrix} A \\ D \end{bmatrix} \quad (4)$$

Here, $A$ and $D$ are the complex amplitude of incident wave whereas $B$ and $C$ are the complex amplitude of scattered wave from the left and the right sides of the cavity respectively. The physical solutions of equation (4) under open boundary condition are associated with both real and complex values of $\omega$. A pole of the $S$-matrix which appears in the complex frequency plane, physically indicates that the scattering resonances correspond to purely outgoing boundary conditions on the $S$-matrix ($A = D = 0$) occur. In a lossless cavity the appearance of equidistant poles of the $S$-matrix are required to obey current conservation and causality. Hence the poles occur only at complex values of k in the lower half plane i.e. with a negative imaginary part.

### IV. OPTICAL PROPERTIES AND PERFORMANCES OF INTERACTING RESONANCES

To realize the presence of an avoided crossing in the eigenvalue spectrum between the interacting poles, we plot the real and imaginary components of eigen values of $S$-matrix. In presence gain/ loss the real part of resonance wave numbers define frequencies at which linear scattering is resonantly enhanced and the imaginary parts represent the out-coupling loss. In a passive cavity, we deliberately locate two poles of $S$ matrix in the lower half of the complex frequency plane for a certain frequency limit whose real value lies between 8.3 to 8.7 (in $\mu m^{-1}$) and imaginary part lies between $-1$ to $+1$. Accordingly, when gain is added non-uniformly the chosen poles of the passive cavity are now mutually coupled. With the increasing amount of gain, we observe the state repulsion in the eigenvalue plot of the cavity as shown in Figs. 2(a)-(b). Accordingly, for a given set of $(\gamma, \tau)$ we have identified the regime where an avoided level crossing exists in the complex $k$-plane. In the case of coupled bound states where the underlying Hamiltonian is hermitian and off-diagonal coupling is present between the states, there exist the level repulsion phenomena. So, the extension of repulsion effect is quite natural from bound to leaky states. However, the fact that leaky/resonant states are having complex eigen values, opens up a rich scenario of crossing and anti-crossing of its real and imaginary parts for the coupled states. The term crossing and anti-crossing refer here to the situation that the two real or imaginary parts of the eigenvalues are calculated as a function of $\gamma$ and $\tau$. Crossing is observed when for a given set of $\gamma_c$ and $\tau_c$ value the real/imaginary parts of $k$'s become equal i.e. $\Re(k1(\gamma_c,\tau_c)) = \Re(k2(\gamma_c,\tau_c))$ where $k1$ and $k2$ are the wave numbers of the coupled states respectively. Besides, when for all values of $\gamma$ and $\tau$ the real/imaginary components of $k$'s are different, there exists an anti-crossing. In FIG. 2, for $\tau$ set at 0.274, we have shown the crossing behavior of $\Re(k)$ and anti-crossing behavior of $\Im(k)$. Alternatively, the opposite scenario is observed at smaller $\tau$ values; i.e. anti-crossing behavior of $\Re(k)$ and crossing behavior of $\Im(k)$ respectively. This scenario of first kind associated with the $ARC$ between the chosen states near the transition regime of $\tau$ (set at 0.274) has been illustrated in Fig. 2(c)-(d). Near the $\Re(k) = 8.48\mu m^{-1}$, interacting modes cross at $\gamma = 0.0875$ i.e. point of virtual frequency crossing (as shown in Fig. 2(c)). Choice of this specific value of $\tau$, which marks a special point in the $2D$ real parameter space i.e. $(\gamma, \tau)$ system parameter space was driven by the two key factors. Primarily motivated by the fact that this kind of $ARC$ favors formation of longer-lived state, and hence lifetime is maximized for the longer lived state at this point of the parameter space for the interacting states. Moreover the required $\gamma$ value is minimum for the optimized $Q$-factor enhancement of the corresponding longer-lived state. These will be discussed in the following sections in more detail. Thus occurrence of such special $ARC$ in our suitably designed cavity will eventually turn on some unconventional optical properties for potential technological applications in integrated photonics. The eigen-functions corresponding to the two coupled states are plotted near the special avoided crossing regime and depicted in FIG. 3. The exchange of states is clearly displayed in the squared eigen-function plot with slowly increasing $\gamma$ parameter.

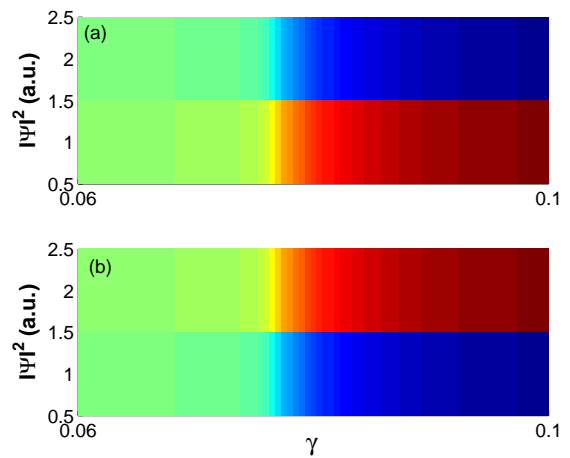

FIG. 3: (Color online) Variation of the cavity eigenvectors (squared in arbitrary unit) with increasing gain around the frequency $ARC$ are shown in (a) & (b) respectively.

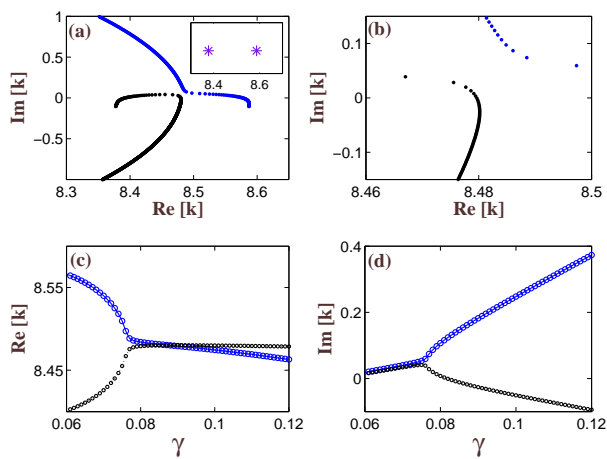

FIG. 2: (Color online) (a) Trajectories of two interacting poles due to addition of non-uniform homogeneous gain/loss in the cavity which shows $ARC$ near $8.485\mu m^{-1}$ of real value of frequency. Inset: two arbitrarily chosen poles in the passive cavity for coupling. (b) Magnified view of the $ARC$ is clearly evident. (c) & (d) Variation of real and imaginary part of frequencies respectively with gain. $ARC$ is associated with a crossing in real part and anti-crossing in imaginary part between the interacting states.

Manipulation of lifetime of resonances via mutual coupling between interacting states and eventually to produce longer lived state with significant enhancement in $\tau_L$ is primarily aimed in this paper. The numerically estimated $Q$-values with increasing gain for a given $\tau = 0.274$ is plotted in FIG. 4 (a). Clearly the figure shows an anti-crossing phenomenon of $Q$-variation. The control of cavity emission around this regime has become an important application of micro-cavities. Near the frequency $ARC$, the $Q$-value of the longer-lived state (in blue solid line) is enhanced in more than two orders of magnitude with respect to the passive cavity mode and obtained the value $\approx 0.5 \times 10^4$. While the $Q$-value of the counterpart shorter-lived state (in black dotted line) drastically loses its quality and $Q$ is reduced to 30. The peak of $Q$-variation for the longer lived state is positioned at $\gamma = 0.082$. The $Q$-value variation for resonance between two identical resonant modes can be described by their parity symmetry dependence. When the resonant modes with odd parity are excited then intensity is suppressed by destructive interference resulting overall increase in $Q$-factor that indicates less light leakage. Whereas if the resonant modes with even parity are excited then intensity is enhanced by constructive interference and consequently $Q$-factor is decreased [41].

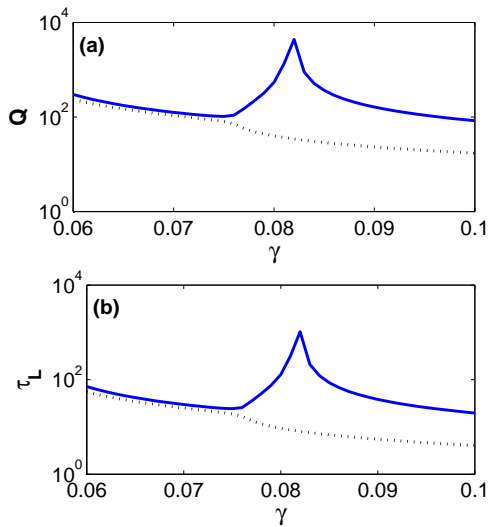

FIG. 4: (Color online) 4. (a) & (b) $Q$ values and life time $\tau_L$ of individual states as a function of increasing gain (Blue solid lines for longer lived states and black dotted lines for shorter lived state) for a chosen $\tau$ value of 0.274.

Moreover, we show the formation of longer lived state and shorter lived state due to interaction between modes and mutual exchange of imaginary part of $k$ by plotting life times as a function of gain. Cavity lifetime plot for the coupled states corresponding to Fig. 4(a) is exhibited in Fig. 4(b). Life time of longer lived state is obtained as much as 128 times higher than that of shorter lived state. Hence, the enhancement in quality factor is also nearly 128 times for the longer lived state than that of shorter lived one. Evidently, in reality the overall cavity performance will be governed by the high-$Q$ state. Once explored for low threshold lasing, this state would eventually favour efficient coupling to other targeted guiding optical components. To note, the maximum achievable $Q$-value for the longer lived state, for a chosen cavity size and the corresponding required gain, can be optimized with respect to the strength of coupling, selection of frequencies of the interacting resonances i.e. choice of suitable set of coupled states and nature of the associated $ARC$ in the operating regime. Thus these cavity enhancement factors can be exploited to realize threshold control of microcavity lasers. More recently, several groups have reported impressive concerted optimization of mode volume, $Q$-factor and lifetime in such micro-cavities. However, synthesis of excess noise reduction has not been performed. Rather with no obvious reason it is always mentioned in separate. The *Petermann $K$-factor* is the measure of non orthogonality of the coupled states, serves as an efficient tool to obtain additional information when such open system is externally probed. Hence, we have calculated $K$-factor over the entire regime ($\gamma \to 0.06$ to $0.1$) of $Q$-anti crossing and displayed in FIG. 5. Owing to the mode-mode interaction around the $ARC$, the $K$- factor increases inducing instabilities and obtains a peak for both the states at $\gamma = 0.076$. Interestingly, the position of the maximum value of $K$ does not coincide with the highest $Q$-value in the parameter space ($\gamma = 0.082, \tau = 0.274$). Rather with suitable choice of operation in the parameter space, results in minimization of $K$-factor ($\approx 1.86$) near the $Q$-factor peak at $\gamma = 0.082$ of the longer-lived state. It can be appreciated that this $K$ value is sufficiently close to the ideal value of 1. It is also interesting to note from Fig. 5 that $K$-factor is enhanced for both the states (blue solid curve for longer lived state and black dotted curve for shorter lived state respectively) irrespective of its degradation or enhancement of quality. Moreover, the $K$-peaks are located near the $\gamma$ value 0.076. Beyond this point, even if we increase the pumping strength ($\gamma$), the noise factor decreases for both the state. The rate of fall from the peak value of $K$ with increasing $\gamma$ is faster as evident from the asymmetry in $K$-behavior around $\gamma = 0.076$. This is essentially due to the fact that the interacting modes depart with faster speed after reaching the distance of closest approach adjacent to the virtual crossing point (as shown in the Fig. 2(a)). Particularly, the variation of $K$-factor results from the inherent mode properties of the near-degenerate cavity. Thus our designed degenerate cavity would support ultrahigh-$Q$ state with simultaneous excess noise reduction. In other words, in addition to the robustness to ultra-high-$Q$ micro resonators, our study would introduce a new avenue towards stabilization of outputs from such systems. The characteristic response of the cavity near another transition point at $\tau = 0.46$ associated with a special type $ARC$ is displayed as a specimen plot in FIG. 6. Even though $Q$-enhancement is achieved in this case near the $Q$-crossing associated with the frequency $ARC$, however the peak values of $K$-variation almost coincide with $Q$-peaks in the $\gamma$ axis. The cavity-configuration and pump-size dependency of the $K$-factor near the degeneracies has been investigated in [42] in the context of laser instabilities.

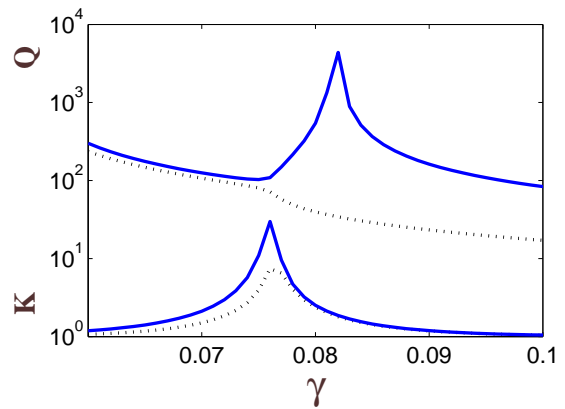

FIG. 5: (Color online) The $K$ factors of the interacting states as a function of increasing gain near the $Q$-avoided crossing. Blue solid line represents longer lived state and black dotted line for shorter lived state respectively.

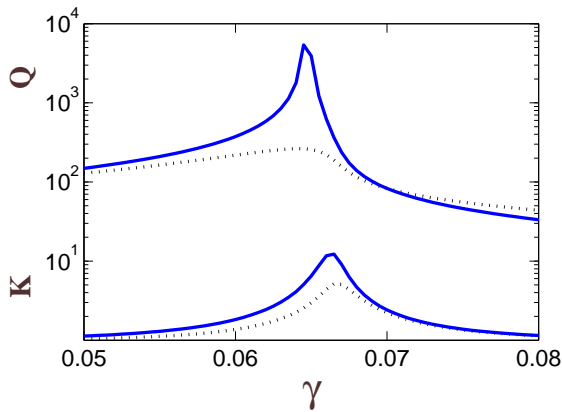

FIG. 6: (Color online) The $K$ factors of the interacting states as a function of increasing gain near the $Q$-crossing for chosen $\tau$=0.46. Blue solid line represents longer lived state and black dotted line for shorter lived state respectively.

From Figs. 2(c), 4(a) and 5, it is established that the position of $Q$-peak is near $\gamma = 0.082$ which is almost equidistant from the location of maximum $K$- at $\gamma = 0.076$ and crossing point of $\Re(k)$ at $\gamma = 0.0875$ respectively. Transmission behavior of the cavity over this interesting region is shown in FIG. 7 and evidently favors the high-$Q$ longer lived mode near $\gamma = 0.082$. Unlike the critical implementation of excellent surface finish, deformation free resonators, simply carefully chosen partial pumping strength near the degeneracies would result in a stable quality-cavity propagation effects exploiting state-of-the-art fabrication techniques. Hence, performance control is less susceptible to other fabrication tolerance related parameters. The bulk optical loss of the chosen material is already exceptionally low to host the high-$Q$ performance. Further improvements in terms of the reported $Q$-enhancement factor along with increase in life time and order of magnitude reduction in $K$- is achievable with further optimization and scaling of the cavity with proper choice of operating $k$.

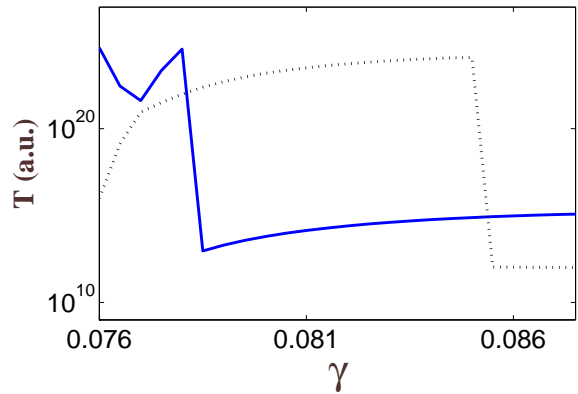

FIG. 7: (Color online)Transmission coefficients of the interacting states as a function of increasing gain near the $Q$-anti crossing for chosen $\tau = 0.274$. Blue solid line represents longer lived state and black dotted line for shorter lived state respectively.

## V. CONCLUSION

In this study we have shown, how with appropriate tuning of the pumping strength in a nonuniformly pumped simple gain/loss cavity we can minimize the excess noise while achieving ultra-high quality factor for longer lived state with adequately scaled-up lifetime. The decay rate for the state with highly reduced $Q$ is consequentially speeded up. Our deliberate choice of a partially pumped cavity with two coupled states is proven to be ideal to study excess noise in such configurations. More importantly, substantial improvement in the propagation characteristics of the targeted state for cavity emission is achieved with improved stability in performance in terms of sufficiently suppressed (close to ideal value of 1) unwanted but omni-present *Petermann K*-noise factor. Our proposed scheme should open up a new avenue to design high performance low threshold photonic devices based on such microcavities. With the state-of-the-art micro fabrication techniques, we expect implementation of this type of noise suppressed low-loss cavity designs is feasible on-chip.

S. N. Ghosh acknowledges the financial support by department of science and technology, India as a INSPIRE Faculty Fellow [IFA-12; PH-13].